\documentclass[11pt,a4paper]{article}

\usepackage{amsmath,amssymb}
\usepackage{amsfonts}
\usepackage{eucal}

\usepackage{xy}
\xyoption{all}

\title{%
GEOMETRIC HAMILTON--JACOBI THEORY\\
FOR NONHOLONOMIC DYNAMICAL SYSTEMS}
\author{\sc Jos\'e F. Cari\~nena\thanks{{\bf e}-{\it mail}: jfc@unizar.es}\\
\tabaddress{Departamento de F\'{\i}sica Te\'orica\\
Facultad de Ciencias,
Universidad de Zaragoza,
50009 Zaragoza. Spain}
\\
{\sc Xavier Gr\`acia\thanks{{\bf e}-{\it mail}: xgracia@ma4.upc.edu}}\\
 \tabaddress{Departament de Matem\`atica Aplicada IV,
Universitat Polit\`ecnica de Catalunya\\
Campus Nord UPC edifici~C3,
C. Jordi Girona~1, 08034 Barcelona, Catalonia, Spain}
\\
{\sc Giuseppe Marmo\thanks{{\bf e}-{\it mail}: giuseppe.marmo@na.infn.it}}\\
 \tabaddress{Dipartimento  di Scienze Fisiche,
Universit\'a Federico II di Napoli, and INFN, Sezione di Napoli\\
Complesso Univ.\ di Monte Sant'Angelo, Via Cintia,
80126 Napoli. Italy}
\\
{\sc Eduardo Mart\'{\i}nez\thanks{{\bf e}-{\it mail}: emf@unizar.es}}\\
 \tabaddress{Departamento de Matem\'atica Aplicada and IUMA\\
Facultad de Ciencias, Universidad de Zaragoza, 50009 Zaragoza. Spain}
\\
{\sc Miguel C. Mu\~noz--Lecanda\thanks{{\bf e}-{\it mail}:
 matmcml@ma4.upc.edu}},
{\sc Narciso Rom\'an--Roy\thanks{{\bf e}-{\it mail}: nrr@ma4.upc.edu}}\\
 \tabaddress{Departamento de Matem\'atica Aplicada 4\\
  Edificio C-3, Campus Norte UPC.
  C/ Jordi Girona 1. E-08034 Barcelona. Spain}}

\date{14 August 2009}
\def\tabaddress#1{{\small\it\begin{tabular}[t]{c}#1
\\[1.2ex]\end{tabular}}}
\def\R{\mathbb{R}}

\def\pd#1#2{\frac{\partial #1}{\partial#2}}

\def\<#1>{\langle#1\rangle}
\def\Ker{\mathop{\rm Ker}\nolimits}
\def\derpar#1#2{\frac{\partial{#1}}{\partial{#2}}}
\newcommand{\vectorfields}[1]{\mathfrak{X}(#1)}
\newcommand{\Sec}[2][]{\operatorname{Sec}\nolimits_{#1}(#2)}
\let\sec\Sec
\newcommand{\cinfty}[1]{C^\infty(#1)}

\newcommand{\Ver}[1]{\operatorname{Ver}(#1)}

\newcommand{\nhbr}[2]{\{#1,#2\}^{\operatorname{nh}}}

\newcommand{\D}[1][]{\mathcal{D}_{#1}}
\newcommand{\Do}[1][]{\mathcal{D}_{#1}^\circ}
\newcommand{\tDo}[1][]{\widetilde{\mathcal{D}_{#1}^\circ}}
\newcommand{\GL}[1][]{G^{L}_{#1}}
\newcommand{\GLD}[1][]{G^{L\mathcal{D}}_{#1}}
\newcommand{\TDD}[1][]{\CMcal{T}^{\mathcal{D}}_{#1}\mathcal{D}}
\newcommand{\wLD}[1][]{\omega^{L\mathcal{D}}_{#1}}
\newcommand{\eLD}[1][]{\varepsilon^{L\mathcal{D}}_{#1}}

\def\transp#1{{}^{t}\kern-.15em\relax#1}
\def\tanvec#1{{\partial \over \partial #1}}
\def\restric#1#2{{\left. #1 \right|_{#2}}}
\newtheorem{theorem}{Theorem}

\newtheorem{proposition}{Proposition}
\newtheorem{definition}{Definition}

\newtheorem{statement}{Statement}
\newtheorem{remarkth}{Remark}
\newenvironment{remark}{\begin{remarkth}\upshape}{\end{remarkth}}
\def\bit{\begin{itemize}}
\def\eit{\end{itemize}}
\def\bea{\begin{eqnarray}}
\def\eea{\end{eqnarray}}
\def\beann{\begin{eqnarray*}}
\def\eeann{\end{eqnarray*}}
\def\ben{\begin{enumerate}}
\def\een{\end{enumerate}}
\def\qed{\ifvmode\removelastskip\fi
{\unskip\nobreak\hfil\penalty50\hbox{}\nobreak\hfil \hbox{\vrule
height1.2ex width1.2ex}\parfillskip=0pt \finalhyphendemerits=0
\par\smallskip}}
\def\Tan{T}
\def\d{d}

\let\ds=\displaystyle

\def\Lie{\mathop{{\cal L}\strut}\nolimits}

\newcommand{\map}[3]{#1\colon#2\to#3}
\newcommand{\Real}{\mathbb{R}}

\def\texthook{\vrule height 0pt depth 0.4pt width 3.5pt
          \vrule height 5pt depth 0.4pt \kern 3pt}
\def\scripthook{\vrule height 0pt depth 0.2pt width 1.5pt
                \vrule height 3pt depth 0.2pt width 0.2pt \kern 1pt}

\newcommand{\sode}{\textsc{sode}}

\newcommand{\set}[2]{\left\{\,#1\left.\vphantom{#1#2}\,\right\vert\,#2\,\right\}} %set

\newcommand{\pai}[2]{\langle#1,#2\rangle}

\let\vf\vectorfields

\newcommand{\at}[1]{\Big|_{#1}}
\begin{document}
%%%%%%%%%%%%%%%%%%%%%%%%%%%%%%%%%%%%%%%%%%%%%%%%%%%%%%%%%%%%%%%%

\maketitle

\thispagestyle{empty}

\begin{abstract}
\noindent
The geometric formulation of Hamilton--Jacobi theory for systems with 
nonholonomic constraints is developed, 
following the ideas of the authors in previous papers. 
The relation between the solutions of the Hamilton--Jacobi problem 
with the symplectic structure defined from the Lagrangian function and the constraints 
is studied. 
The concept of complete solutions and their relationship with constants of motion, 
are also studied in detail. 
Local expressions using quasivelocities are provided. 
As an example, the nonholonomic free particle is considered.
\end{abstract}



\noindent
{\sl PACS number (2003)}:
02.40.Yy, 45.20.Jj

%%%%%%%%%%%%%%%%%%%%%%%%%%%%%%%%%%%%%%%%%%%%
\clearpage
%%%%%%%%%%%%%%%%%%%%%%%%%%%%%%%%%%%%%%%%%%%%%%
% \tableofcontents
%%%%%%%%%%%%%%%%%%%%%%%%%%%%%%%%%%%%%%%%%%%%%%%

%%%%%%%%%%%%%%%%%%%%%%%%%%%%%%%%%%%%%%%%%%%%%%%%%%%%%%%%%%%%%%%%
%%%%%%%%%%%%%%%%%%%%%%%%%%%%%%%%%%%%%%%%%%%%%%%%%%%%%%%%%%%%%%%%
\section{Introduction}
%%%%%%%%%%%%%%%%%%%%%%%%%%%%%%%%%%%%%%%%%%%%%%%%%%%%%%%%%%%%%%%%
%%%%%%%%%%%%%%%%%%%%%%%%%%%%%%%%%%%%%%%%%%%%%%%%%%%%%%%%%%%%%%%%

In classical mechanics, Hamilton--Jacobi theory tries to integrate 
a Hamiltonian system of differential equations
through an appropriate canonical transformation
\cite{Arn,JS}.
The equation to be satisfied by the generating function of this transformation
is a partial differential equation,
and having enough solutions to it finally leads to the integration
of the system.
The Hamilton--Jacobi equation is also very close,
from the classical side,
to the Schr\"odinger equation of quantum mechanics
---see for instance \cite{MMMcq}.
For these reasons, 
Hamilton--Jacobi theory has been a matter of continuous interest.

From the viewpoint of geometric mechanics,
the intrinsic formulation of Hamilton--Jacobi equation is also
clear
\cite{Ab,LM-87,MMM}.
Nevertheless, in a recent paper
\cite{HJteam} 
we presented a new geometric framework for the Hamilton--Jacobi theory.
The motivation for this work was that the usual formulation of the
Hamilton--Jacobi equation heavily relies on the symplectic structure
of the phase space.
However, there are interesting integrable systems 
that have alternative Lagrangian (and Hamiltonian) formulations;
two different Lagrangians for the same dynamics may lead to
two different symplectic structures,
and therefore one may wonder about the relevance of a concrete 
symplectic structure and its relation with the solutions of the
Hamilton--Jacobi problem.
Following this program,
we formulated the Hamilton--Jacobi equation
both in the Lagrangian and in the Hamiltonian formalisms
of time-independent mechanics,
and studied the relations between the solutions of the Hamilton--Jacobi
equation and the symplectic form;
we recovered the usual Hamilton--Jacobi equation as a special case
in our generalised framework.
Additional details on the relationship between Hamilton--Jacobi equation
and the geometric structures of mechanics have recently been presented in
\cite{HJteam-K}.

Within the Lagrangian formulation,
dynamics is described by a second-order vector field
$\Gamma$ defined on the tangent bundle $TQ$ 
of the configuration manifold~$Q$.
The first step in our formulation is 
to describe the integral curves of~$\Gamma$
as the canonical liftings of the integral curves of 
a family of vector fields $X_\lambda$ on~$Q$.
From a geometrical viewpoint, this is pretty simple:
each of these vector fields has to be $X_\lambda$-related to~$\Gamma$.
The usual formulation of Hamilton--Jacobi equation corresponds to the case
where the image of $X_\lambda$ is a Lagrangian submanifold of $TQ$
with respect to the symplectic form $\omega_L$.
With some changes, the same formulation can be given in the Hamiltonian
framework;
properly speaking, it is in this case that we recover the usual
Hamilton--Jacobi theory.

Our work 
\cite{HJteam} 
was mainly devoted to regular autonomous Lagrangians.
However, we also considered the time-dependent case (through the
so-called homogeneous formalism) 
as well as a special instance of singular Lagrangians:
those not yielding Lagrangian constraints.
It was clear that more general situations could be given a
similar description, and it is the purpose of this paper to
consider the very important issue of 
mechanical systems with nonholonomic constraints
---that is, non-integrable constraints depending on the velocities.

Nonholonomic mechanical systems have been discussed since long ago.
There are many papers dealing with geometric aspects of such systems,
beginning with
\cite{VF-diff},
and including different viewpoints as
\cite{BS-nonh,GM-nh,ILLM,KM-rel,LM-gnh,Mar-approaches,MVB-nonh}
---see also
\cite{BKMM-sym,CL-redrec,Koi-red,MeLa,Sn98}.
When a nonholonomic system is regular,
at the end, there is a well-defined dynamics on the submanifold
$\D \subset TQ$ defined by the constraints. 
Therefore, it seems quite straightforward to apply
our previous framework developed in \cite{HJteam}
for the Hamilton--Jacobi theory to the case of
nonholonomic mechanical systems,
and, in fact, this has been done in some recent papers as 
\cite{leones1,leones2,blo},
where the Hamiltonian case and some applications are analyzed in deep,
as well as in other instances like classical field theories 
\cite{leones3}.
Note, however, that the relation with the symplectic structure is not
so much clear, and this is one of the points we address
in the present paper,
where this new geometric perspective for the Hamilton--Jacobi problem
is performed under the Lagrangian formalism.
In this sense, our approach could be considered as complementary to that
developed in 
\cite{leones1,leones2,blo}.
As in our previous work  
\cite{HJteam}, 
we state the standard classical nonholonomic Hamilton--Jacobi problem
as a particular case of a more general one.
Furthermore, we consider two Lagrangian frameworks for this:
a plain formulation on the velocity space and also
an intrinsic formulation on the constraint submanifold
(the so-called distributional approach to nonholonomic mechanics).
Finally, in the same lines of our previous paper, 
we discuss complete solutions for the Hamilton--Jacobi problem 
and their relationship with constants of motion.

\medskip

The paper is organised as follows.
In section~2 we give a short account of nonholonomic mechanics.
The Hamilton--Jacobi problem for Lagrangian nonholonomic systems is
presented in section~3 in both the general and the restricted (standard) versions.
Section~4 is devoted to the study of the same problem in an
intrinsic formulation.
% , which allows to justify the statement of the
% restricted statement of the problem. 
Local coordinate expressions are given in section~5 by using quasivelocities. 
Complete solutions are studied in section~6.
Finally, a detailed example, the nonholonomic free particle,
is presented in section~7.

%%%%%%%%%%%%%%%%%%%%%%%%%%%%%%%%%%%%%%%%%%%%%%%%%%%%%%%%%%%%%%%%
%%%%%%%%%%%%%%%%%%%%%%%%%%%%%%%%%%%%%%%%%%%%%%%%%%%%%%%%%%%%%%%%
\section{Nonholonomic  Lagrangian systems}
%%%%%%%%%%%%%%%%%%%%%%%%%%%%%%%%%%%%%%%%%%%%%%%%%%%%%%%%%%%%%%%%
%%%%%%%%%%%%%%%%%%%%%%%%%%%%%%%%%%%%%%%%%%%%%%%%%%%%%%%%%%%%%%%%

%\subsection{Lagrangian formalism for nonholonomic systems}

We consider an $n$-dimensional manifold $Q$, 
its tangent bundle $\map{\tau_Q}{TQ}{Q}$, and 
a constraint submanifold, which we assume to be a
vector subbundle $\D\subset TQ$ of rank~$r$. 
We consider the annihilator
$\Do\subset T^*Q$ and the set $\tDo\subset T^*(TQ)$
defined by $\tDo=\set{\alpha\circ T\tau_Q\in T^*(TQ)}{\alpha\in\Do}$;
this is a vector bundle over $TQ$, whose fibre at a point $v \in TQ$,
such that $\tau_Q(v)=q$,  
is more explicitly described as
\[
\tDo[v] =
\set{\lambda_v \in T^*_v(TQ)}
{\text{there exists $\alpha_q \in \Do[q]$ such that 
$\lambda_v = \alpha_q \circ T_v\tau_Q$}}.
\]

Given a Lagrangian function $L\in\cinfty{TQ}$,
we consider the nonholonomic system defined by the Lagrangian $L$
and the linear constraints given by $\D$, that is, 
only velocities in $\D$ are admissible.
The Lagrange--d'Alembert principle states that the dynamics
of the system is given by the integral curves (with initial condition in $\D$) 
of the vector fields
$\Gamma\in\vf{TQ}$ tangent to $\D$
that satisfy the second-order condition and
the Lagrange--d'Alembert equation 
(see for instance \cite{LM-gnh})
\begin{equation}
\label{Lagrange-D'Alembert.external}
(i_\Gamma\omega_L-dE_L)\vert_{\D} \in \Sec{\tDo} \ ,
\end{equation}
where $\omega_L$ is the Lagrange 2-form associated with~$L$.
This expression means that, on the points of $\D$, the 1-form
$i_\Gamma\omega_L-dE_L$ takes its values in the codistribution $\tDo$.

From now on we assume that $L$ is a regular Lagrangian,
which means either that
its fibre derivative (Legendre transformation)
$\mathcal{F}L \colon TQ \to T^*Q$ is a local diffeomorphism,
that the Lagrange 2-form $\omega_L$ is a symplectic form,
or that
its fibre Hessian $\mathcal{F}^2L = \GL[] \colon TQ \to T^*Q \otimes T^*Q$
is everywhere a nondegenerate bilinear form.
Given $u, v, w \in T_qQ$,
the fibre Hessian of the Lagrangian can also be expressed as
$\GL[u](v,w)=\omega_L(\tilde{v},w^V_u)$,
where $\tilde{v} \in T_uTQ$ is any vector which projects onto~$v$,
and $w^V_u$ is the vertical lift of $w$ on the point $u$.

The nonholonomic system $(L,\D)$ is said to be {\sl regular}
if there is a unique solution to Lagrange--d'Alembert equation.
Here uniqueness must be understood as follows:
two solutions are considered equal if they coincide when 
restricted to~$\D$.
%%%%%%%%%%%%%%%%%%%%%%%%%%%%%%%%%%%%%%%
% In this case, $\Gamma$ is tangent to $\D$, and the dynamics is consistent;
%otherwise, the tangency condition must be imposed.
%%%%%%%%%%%%%%%%%%%%%%%%%%%%%%%%%%%%%%%%%%

There are several equivalent ways to ensure regularity of 
the constrained system.
%If $\widehat\omega_L \colon \Tan(\Tan Q) \to \Tan^*(\Tan Q)$
%is the isomorphism defined by~$\omega_L$,
%we define the vector subbundle $F$ of $\Tan(\Tan Q)|_{\D}$ as
%$$
%F = \widehat\omega_L^{-1} (\tDo) \ .
%$$
%We also define the subundle $\TDD\subset T\D\,\to\D$ by
We define the subundle $\TDD\subset T\D\,\to\D$ by
\[
\TDD=\set{V\in T\D}{T\tau_Q(V)\in\D} \ .
\]
We also consider the restriction $\GLD$ of the fibre Hessian $\GL$ 
to the distribution $\D$. 
Then (see for instance \cite{nonholoid}):
\begin{theorem}
\label{regularity}
The following properties are equivalent:
\begin{enumerate}
\item The constrained Lagrangian system $(L,\D)$ is regular,
\item $\Ker\GLD=\{0\}$.
%\item $TT Q|_{\D}=T\D\oplus F$.
\item $TT Q|_{\D}=\TDD\oplus(\TDD)^\perp$,
\end{enumerate}
where $(\TDD)^\perp$ denotes the orthogonal complement of
$\TDD$ with respect to the symplectic form~$\omega_L$.
\end{theorem}

In the regular case, the constrained dynamics
can be found by projection of the free dynamics according
to the decomposition given in item~3.  
It follows that the dynamical vector field is a \sode\ on~$\D$,
that is, $\Gamma$ is tangent to $\D$ and 
$T\tau_Q(\Gamma(v))=v$ for every $v\in\D$.

%%%%%%%%%%%%%%%%%%%%%%%%%%%%%%%%%%%%%%%%%%%%%%%%%%%%%%%%%%%%%%%%
%%%%%%%%%%%%%%%%%%%%%%%%%%%%%%%%%%%%%%%%%%%%%%%%%%%%%%%%%%%%%%%%
\section{The Lagrangian Hamilton--Jacobi problem for honholonomic systems}
%%%%%%%%%%%%%%%%%%%%%%%%%%%%%%%%%%%%%%%%%%%%%%%%%%%%%%%%%%%%%%%%
%%%%%%%%%%%%%%%%%%%%%%%%%%%%%%%%%%%%%%%%%%%%%%%%%%%%%%%%%%%%%%%%
\label{lfnhs}

As in our previous paper
\cite{HJteam},
we decompose the study of the Hamilton--Jacobi problem
for a nonholonomic Lagrangian system
in two pieces:
first, we consider a general setting to describe the solutions
of the nonholonomic dynamics $\Gamma$ on~$\D$
in terms of the solutions of a family of 
first-order differential equations;
second,
we study the interplay of these first-order vector fields
with the corresponding symplectic structure,
and impose additional conditions on them
in order to simplify the problem.
All this is performed in the Lagrangian formalism
---the case of Hamiltonian formalism can be developed in quite a
similar way.

%%%%%%%%%%%%%%%%%%%%%%%%%%%%%%%%%%%%%%%%%%%%%%%%%%%%%%%%%%%%%%%%
\subsection{General Lagrangian nonholonomic Hamilton--Jacobi problem}
%%%%%%%%%%%%%%%%%%%%%%%%%%%%%%%%%%%%%%%%%%%%%%%%%%%%%%%%%%%%%%%%

Following the same lines as in 
\cite{HJteam},
we formulate the Hamilton--Jacobi problem in this way:

\begin{statement}
{\bf (General Lagrangian nonholonomic Hamilton--Jacobi problem)}
Given a regular nonholonomic Lagrangian system $(L,\D)$,
with dynamics given by a \sode\ vector field $\Gamma\in\vectorfields{\D}$,
the {\rm general Lagrangian nonholonomic Hamilton--Jacobi problem}
consists in finding the vector fields $X\colon Q\to TQ$
such that, if $\gamma\colon\R\to Q$ is an integral curve of $X$,
then $\dot\gamma \colon \R \to TQ$
takes values in $\D \subset TQ$ and it
is an integral curve of $\Gamma$; that is,
\[
X\circ\gamma=\dot\gamma \ \Longrightarrow \
\Gamma\circ\dot\gamma=\dot{\overline{X\circ\gamma}} 
\quad
\hbox{and}
\quad  
\hbox{$\dot\gamma(t)\in\D$ for each $t\in\R$} \, .
\]
Any of such $X$ is said to be a 
{\rm solution to the general Lagrangian nonholonomic Hamilton--Jacobi problem}.
\end{statement}

\begin{theorem}
A vector field $X \in \vectorfields{Q}$ 
is a solution to the general Lagrangian nonholonomic
Hamilton--Jacobi problem if, and only if,
$X\in\Sec{\D}$ and $\Gamma\circ X=TX\circ X$.
\label{t2}
\end{theorem}
\begin{proof}
% It is a consequence of Proposition~\ref{prop-related}.
Let $X\in\vf{Q}$ be a solution to the general nonholonomic Hamilton--Jacobi problem.
For every $q\in Q$, let $\gamma$ be the integral curve of $X$ starting at $q$;
that is, $\dot{\gamma}=X\circ\gamma$ and $\gamma(0)=q$.
Then $\eta=\dot{\gamma}$ is a solution to the constrained problem;
that is, $\eta(0)\in\D$ and $\dot{\eta}=\Gamma\circ\eta$.
From the first one we have that
$X(q)=X(\gamma(0))=\dot{\gamma}(0)=\eta(0)\in\D$.
As $q$ is arbitrary, it follows that $X$ takes values in $\D$. 
Moreover,
\begin{align*}
(\Gamma\circ X)(q)
&=(\Gamma\circ X\circ\gamma)(0)
=(\Gamma\circ\eta)(0)
=\dot{\eta}(0)
=\frac{d\dot{\gamma}}{dt}(0)
=\frac{d}{dt}(X\circ\gamma)(0)\\
&=(TX\circ\dot{\gamma})(0)
=(TX\circ X\circ \gamma)(0)
=(TX\circ X)(q) ,
\end{align*}
from which it follows that $\Gamma\circ X=TX\circ X$.

Conversely, let $X$ be a vector field taking values in $\D$ such that
$\Gamma\circ X=TX\circ X$.
If $\gamma$ is an integral curve of $X$
then  $\eta=X\circ\gamma$ is an integral curve of $\Gamma$:
\[
\Gamma\circ\eta
=\Gamma\circ X\circ\gamma
=TX\circ X\circ\gamma
=TX\circ \dot{\gamma}
=\frac{d}{dt}(X\circ\gamma)
=\dot{\eta}.
\]
In addition, as $\eta(0)=X(\gamma(0))\in\D$, 
it follows that $\eta$ starts at $\D$,
and hence it is a solution to the constrained dynamics.
\qed
\end{proof}

We can rewrite the above statement as follows:
a vector field $X$ is a solution to the general nonholonomic 
Hamilton--Jacobi problem
if ${\rm Im}\,(X)$ is a submanifold of $\D$ and $\Gamma$ is tangent 
to this submanifold.
Conversely, if $N$ is an $n$-dimensional submanifold of $\D$,
transverse to the fibers and invariant under $\Gamma$,
then locally there exists $X\in\vf{Q}$ such that $N=X(Q)$
and it is a local solution to the general Hamilton--Jacobi problem.

\begin{remark}
As in the unconstrained case (when $\D=TQ$)
the above result can be stated in a more general framework,
and in fact, it can be applied to any vector field on~$\D$
which satisfies the second-order condition.
\end{remark}

The \sode\ $\Gamma$ being the solution of 
the Lagrange--d'Alembert equation (\ref{Lagrange-D'Alembert.external}),
we can take the pullback of such equation by $X$,
and then obtain an equation that does not involve $\Gamma$ explicitly.

\begin{theorem}
\label{thm3}
A vector field $X\in\vf{Q}$ is a solution to the
 general Lagrangian nonholonomic Hamilton--Jacobi problem if, and only if,
 $X\in\Sec{\D}$ and $i_X(X^*\omega_L)-d(X^*E_L)\in\Sec{\Do}$.
\end{theorem}
\begin{proof}
We will use the following preliminary results:
\begin{enumerate}
\item
If $\lambda\in\tDo$, then $\lambda=\alpha\circ T\tau_Q$ for $\alpha\in\Do$,
and we have that $X^*\lambda = \alpha$.
In fact,
\[
\pai{X^*\lambda}{v}=\pai{\alpha\circ T\tau_Q}{TX(v)}=
\pai{\alpha}{T\tau_Q(TX(v))}=\pai{\alpha}{v}
\]
for every $v\in TQ$. 
We will write symbolically this equation as $X^*\tDo=\Do$.

\item
Given a vector field $X\in\vf{Q}$, let $Y$ be the vector field along $X$
defined by $Y=\Gamma\circ X-TX\circ X$. Consider the one-form $\alpha$ in $Q$
given by
$$
\alpha=[X^*(i_\Gamma\omega_L-dE_L)]-[i_X(X^*\omega_L)-d(X^*E_L)] \ .
$$
A straighforward calculation (see \cite{HJteam}) leads to
\[
\alpha_q(v) = \omega_L(X(q)) \,(Y(q),\Tan X(v))
\  ,\qquad q \in Q\ , \ v \in \Tan_qQ \ .
\]
\item
If $X$ is a section of $\D$, then for every $q \in Q$ there exists $v\in\D$
such that $Y(q)=\xi^V(X(q),v)$,
where $\xi^V$ denotes the vertical lift in $\Tan Q$.
Indeed, it is clear that $Y$ take values in the vertical bundle,
so that, for every $q \in Q$ there exists $v\in T_qQ$
such that $Y(q)=\xi^V(X(q),v)$. We have just to prove that $v$ is in $\D$.
On the one hand $\Gamma|_{\D}$ is tangent to $\D$, 
so that $\Gamma\circ X$ takes values in $T\D$, and  on the other hand,
$TX\circ X$ also takes values in $\Tan\D$, therefore we get that 
$Y(q)\in T_{X(q)}\D$. Taking into account that linear constraints for $\D$
are given by the linear functions $\hat{\alpha}$ associated with 1-forms $\alpha$
taking values in $\Do$, we have that, for every $\alpha\in\Sec{\Do}$,
\[
0=Y(q)\hat{\alpha}=\xi^V(X(q),v)\hat{\alpha}=\pai{\alpha_q}{v},
\]
and hence $v\in\D$.
\end{enumerate}

Bearing this in mind, the proof of the theorem is as follows:
\newline
[$\Leftarrow$]
Let $X\in\Sec{\D}$ such that $i_X(X^*\omega_L)-d(X^*E_L) \in \Sec{\Do}$.
As $i_\Gamma\omega_L-dE_L \in \Sec{\tDo}$, 
then $X^*(i_\Gamma\omega_L-dE_L) \in \Sec{\Do}$,
and hence
\[
\alpha=[X^*(i_\Gamma\omega_L-dE_L)]-[i_X(X^*\omega_L)-d(X^*E_L)] \in \Sec{\Do} .
\]
The vector field $Y$ along $X$ is vertical, and at every point is the vertical lift
of an element in $\D$: for every $q \in Q$, there exists $v\in\D$ such that
$Y(q)=\xi^V(X(q),v)$. 
Then for every $w\in\D$ we have that
$$
0=\alpha_q(w)=(\omega_L(X(q))(\xi^V(X(q),v),T_qX(w))=
\GL[X(q)](v,w)=\GLD[X(q)](v,w).
$$
Since this equation holds for every $w\in\D$ and $\GLD$ is regular,
we have that $v=0$, and hence $Y=0$, which proves the statement.

\noindent
[$\Rightarrow$]
If $X\in\Sec{\D}$ and $\Gamma\circ X=TX\circ X$, then $Y=0$,
and hence $\alpha=0$. Therefore
\[
i_X(X^*\omega_L)-d(X^*E_L)=X^*(i_\Gamma\omega_L-dE_L) \in \Sec{\Do}.
\]
This completes the proof.
\qed \end{proof}

%%%%%%%%%%%%%%%%%%%%%%%%%%%%%%%%%%%%%%%%%%%%%%%%%%%%%%%%%%%%%%%%
\subsection{Restricted nonholonomic Lagrangian Hamilton--Jacobi problem}
%%%%%%%%%%%%%%%%%%%%%%%%%%%%%%%%%%%%%%%%%%%%%%%%%%%%%%%%%%%%%%%%

As in the unconstrained case, to solve the generalized Lagrangian nonholonomic
Hamilton--Jacobi problem can be a difficult task; 
thus it is convenient to consider a simplified, and hence less general, problem.

We have seen that $X\in\vf{Q}$ is a solution to the generalized problem
if, and only if, the difference $i_X(X^*\omega_L)-d(X^*E_L)$ takes values in $\Do$. 
So we can look for solutions satisfying that both terms $i_X(X^*\omega_L)$
and $d(X^*E_L)$ are in $\Do$. Furthermore the condition
$i_X(X^*\omega_L) \in \Sec{\Do}$ can be ensured by imposing that
$$
(X^*\omega_L)(\D,\D)=0\ ,
$$
or equivalently $(X^*\omega_L)(\D,\cdot) \in \Sec{\Do}$.
We will plainly say that the restriction of $X^*\omega_L$ to~$\D$ vanishes.

Another possibility could be to impose that $(X^*\omega_L)(\D,\cdot)=0$;
but this is a less general condition.
An additional justification for our choice will be provided in the next section.

In this way, we can state the following restricted
Lagrangian nonholonomic Hamilton--Jacobi problem:

\begin{statement}
{\bf (Restricted Lagrangian nonholonomic Hamilton--Jacobi problem)}
Given a regular nonholonomic Lagrangian system $(L,\D)$,
find those solutions $X$ to the generalized Lagrangian nonholonomic
Hamilton--Jacobi problem such that the restriction of
$X^*\omega_L$ to $\D$ vanishes.
\end{statement}

As a consequence, it follows that 
if $X$ is a solution to the Lagrangian nonholonomic Hamilton--Jacobi problem, 
then $d(X^*E_L) \in \Sec{\Do}$.

\begin{proposition}
A vector field $X\in\vf{Q}$ is a solution to the 
Lagrangian nonholonomic Hamilton--Jacobi problem if, and only if,
\begin{enumerate}
\item 
 $X\in\Sec{\D}$,
\item 
 $(X^*\omega_L)|_{\D}=0$,
\item 
$d(X^*E_L)|_{\D}=0$.
\end{enumerate}
\end{proposition}
\begin{proof}
The direct statement is obvious. 
For the converse, we have that 
$X\in\Sec{\D}$ and both $i_X(X^*\omega_L)$ and $d(X^*E_L)$ 
take their values in $\Do$, 
so that $i_X(X^*\omega_L)-d(X^*E_L) \in \Sec{\Do}$,
and by Theorem~\ref{thm3} the statement holds.
\qed 
\end{proof}

\begin{remark}
It is important to point out that 
every solution to the general (restricted) Lagrangian
Hamilton--Jacobi problem for the unconstrained system
which takes values on $\D$ is, automatically, 
a solution to the general (restricted) Lagrangian 
nonholonomic Hamilton--Jacobi problem.
This may be helpful when looking for solutions 
as we will see in an example later on.
\end{remark}

A particular important case is that of bracket-generating distributions 
(also known as completely nonholonomic distributions). 
A distribution $\D\subset TQ$ is bracket-generating if 
the smallest Lie subalgebra $L_{\D}\subset\vectorfields{Q}$ 
containing $\Sec{\D}$ is the full $\vectorfields{Q}$. 
In other words, we can get a family of vector fields in the distribution $\D$ 
such that every vector $v\in T_qQ$ can be obtained as 
a linear combination of the values at $q$ of such vector fields 
together with repeated brackets. 
In this case we have the following simplification 
(see~\cite{blo}):
   
\begin{proposition}
\label{prop-brac}
Assume that $\D\subset TQ$ is a bracket-generating distribution. 
A vector field $X\in\vf{Q}$ is a solution to the 
Lagrangian nonholonomic Hamilton--Jacobi problem if, and only if,
\begin{enumerate}
\item  $X\in\Sec{\D}$,
\item  $(X^*\omega_L)|_{\D}=0$,
\item  $X^*E_L=\mathrm{constant}$.
\end{enumerate}
\end{proposition}
\begin{proof}
We just have to prove that, for a bracket-generating distribution, 
the condition $d(X^*E_L)\in\sec{\Do}$ is equivalent to $X^*E_L=\mathrm{constant}$.  
The result is true for any function $f$ in $Q$, our case being $f=X^*E_L$. 

Let $f$ be a smooth function on a manifold $Q$ such that $df \in \Sec{\Do}$. 
We first prove that $f$ is constant on the orbits of the family $\mathcal{F}_{\D}$ 
of local vector fields taking values in $\D$. 
Indeed, given a point $q_0$ in the orbit, 
any other point $q_1$ of the orbit is of the form 
$q_1 = (\phi_{t_k}^{X_k} \circ \cdots \circ \phi_{t_1}^{X_1}) (q_0)$ 
for some vector fields $X_i\in\mathcal{F}_D$ and times $t_i\in\Real$. 
Therefore we can get such point by concatenation of a finite number of curves 
of the form $C \,: \, t\in[0,T] \mapsto \phi_t^X(q)$, 
with $q$ a point in the orbit and $X\in\mathcal{F}_D$. 
Integrating $df$ along a curve $C$ of such type 
we get on one hand $\int_Cdf=f(\phi^X_T(q))-f(q)$ and on the other 
$\int_Cdf=\int_0^T\langle df\,,\,X \rangle_{\phi^X_t(q)}\,dt=\int_0^T0\,dt=0$. 
Therefore $f(\phi_T^X(q))=f(q)$ and $f$ is constant along the orbit.

Finally, for a bracket-generating distribution, 
Chow--Rashevsky theorem 
(see for instance
\cite{AS})
ensures that there is only one orbit, the full manifold $Q$. 
Therefore, if $df$ takes its values in~$\Do$, 
then $f$ is a constant function on~$Q$.
\qed
\end{proof}

In the general case, 
provided that the distribution associated with the Lie algebra $L_{\D}$ 
is of constant rank, 
we can restrict our dynamical system to each one of the orbits 
(which are the integral manifolds of the distribution associated 
with $L_{\D}$, 
and hence immersed submanifolds of $Q$),
thus obtaining a Lagrangian system with nonholonomic constraints 
defined by a bracket-generating distribution. 
Hence $X^*E_L$ is constant on every orbit of $L_{\D}$.

%%%%%%%%%%%%%%%%%%%%%%%%%%%%%%%%%%%%%%%%%%%%%%%%%%%%%%%%%%%%%%%%
%%%%%%%%%%%%%%%%%%%%%%%%%%%%%%%%%%%%%%%%%%%%%%%%%%%%%%%%%%%%%%%%
\section{The Hamilton--Jacobi problem in the intrinsic formalism}
%%%%%%%%%%%%%%%%%%%%%%%%%%%%%%%%%%%%%%%%%%%%%%%%%%%%%%%%%%%%%%%%
%%%%%%%%%%%%%%%%%%%%%%%%%%%%%%%%%%%%%%%%%%%%%%%%%%%%%%%%%%%%%%%%

In the above sections we have been using the standard Lagrangian formalism
of nonholonomic constrained problems. Next we develope the theory using
the intrinsic Lagrangian formalism (also called the distributional approach).
This will allow us to justify the choice made for stating the Lagrangian
nonholonomic Hamilton--Jacobi problem.
The distributional approach was initiated by Bocharov and Vinogradov 
\cite{BoVi} 
and further developed by \'Sniatycki and coworkers
\cite{BS-nonh,Sn98}. 
Similar equations, within the more general
framework of Lie algebroids, appear also in~\cite{nonholoid}.

%%%%%%%%%%%%%%%%%%%%%%%%%%%%%%%%%%%%%%%%%%%%%%%%%%%%%%%%%%%%%%%%
\subsection{Intrinsic Lagrangian formalism for nonholonomic systems}
%%%%%%%%%%%%%%%%%%%%%%%%%%%%%%%%%%%%%%%%%%%%%%%%%%%%%%%%%%%%%%%%

In the above standard Lagrangian formalism of nonholonomic
constrained problems, the theory is developed on the whole $TQ$ by
introducing the constraint forces. 
But it is clear that only the values in $\D$ are relevant: 
while the theory depends on the value of the Lagrangian in an open
neighbourhood of $\D$, the final dynamics is defined only on the submanifold~$\D$.
Therefore it is interesting to develop the theory intrinsically in~$\D$.

Recall that we defined the rank $2r$ vector subbundle $\TDD\to\D$ of $T\D$ by
\[
\TDD=\set{V\in T\D}{T\tau_Q(V)\in\D},
\]
and that item 3 in Theorem~\ref{regularity}
expresses the fact that the nonholonomic Lagrangian system is regular
if, and only if, $\TDD$ is a symplectic subbundle of $T(TQ)\vert_{\D}$, 
that is, $\TDD\cap(\TDD)^\perp=\{0\}$. 
Therefore, the restriction $\wLD$
of the symplectic form $\omega_L$ to the subbundle $\TDD$ is regular,
and hence the pair $(\TDD,\wLD)$ is a symplectic vector bundle. 

Similarly, we denote by $\eLD$ the restriction of $dE_L$ to $\TDD$.
It follows that there exists a unique $\Gamma\in\Sec{\TDD}$ such that
\begin{equation}
\label{Lagrange-D'Alembert.internal}
i_\Gamma\wLD=\eLD.
\end{equation}
From the definition of $\wLD$ and $\eLD$
one obtains that the section $\Gamma$ here is
just the restriction to $\D$ of the dynamical vector field $\Gamma$
of the last section, 
and it is a \sode\ in the sense that $T\tau_Q(\Gamma(v))=v$,
for every $v\in\D$. 
We will not make any notational distinction between
the two views of the dynamical vector field.

The advantage of this formulation of the nonholonomic problem
is that we can work entirely in the bundle $\TDD$
following similar arguments to those given for the unconstrained case.
There is only one relevant difference: the 2-form $\wLD$ is not exact.
In fact it even does not make sense to talk about closed forms because
$\TDD$ is not a tangent bundle, neither a Lie algebroid,
except for integrable constraints.

%%%%%%%%%%%%%%%%%%%%%%%%%%%%%%%%%%%%%%%%%%%%%%%%%%%%%%%%%%%%%%%%
\subsection{The general Hamilton--Jacobi problem}
%%%%%%%%%%%%%%%%%%%%%%%%%%%%%%%%%%%%%%%%%%%%%%%%%%%%%%%%%%%%%%%%

In this framework, a solution to the 
\emph{general nonholonomic Lagrangian Hamilton--Jacobi problem}
is a section $\sigma\in\Sec{\D}$ of the vector bundle
$\tau\colon{\D}\to{Q}$ such that the natural lift of its integral curves
are integral curves of $\Gamma$. This statement has sense obviously because
our bundles are subbundles of a tangent bundle, and hence
its sections are vector fields.
It is also clear that this corresponds exactly to the definition in the above section, 
with a change of notation $X\leftrightarrow \sigma$.

Given a section $\sigma\in\Sec{\D}$ we can define the map
 $\CMcal{T}\sigma\colon{\D}\to{\TDD}$ as the restriction of the tangent map
$T\sigma\colon TQ\to T\D$.
% $\CMcal{T}\sigma(v)=T\sigma(v)$ for $v\in\D$. 
It is well-defined since
$T\tau_Q(T\sigma(v))=v\in\D$, so that $\CMcal{T}\sigma(v)\in\TDD$.
With this definition, and according to Theorem~\ref{t2},
a section $\sigma\in\Sec{\D}$ is a solution
to the general nonholonomic Lagrangian Hamilton--Jacobi problem
if, and only if,
\begin{equation}
\label{genintrin}
\Gamma\circ\sigma =\CMcal{T}\sigma\circ\sigma .
\end{equation}

For the following proposition we need a somehow extended notion of the pullback.
 In particular, for a section $\theta$ of the exterior bundle of $(\TDD)^*$,
 we are redefining the meaning of $\sigma^*\theta$
 as the section of the exterior bundle of $\D^*$ given by
\[
(\sigma^*\theta)_q(v_1,\ldots,v_p)=
\theta_{\sigma(q)}\bigl(\CMcal{T}\sigma(v_1),\ldots,\CMcal{T}\sigma(v_p)\bigr),
\]
for $q \in Q$ and $v_1,\ldots, v_p\in \D$.

\begin{proposition}
A section $\sigma\in\Sec{\D}$ is a solution to the general nonholonomic
Lagrangian Hamilton--Jacobi problem if, and only if,
\[
i_\sigma(\sigma^*\wLD)=\sigma^*\eLD.
\]
\end{proposition}
\begin{proof}
% Taking the pullback by $\sigma$ of the equation of motion
% (\ref{Lagrange-D'Alembert.internal}),
% $i_\Gamma\wLD=\eLD$, 
%  we get the equation $i_\sigma(\sigma^*\wLD)=\sigma^*\eLD$.
From equation (\ref{Lagrange-D'Alembert.internal}),
$i_\Gamma\wLD=\eLD$, we have that
 $\sigma^*(i_\Gamma\wLD)=\sigma^*\eLD$.
Now, taking into account that $\Gamma\circ\sigma =\CMcal{T}\sigma\circ\sigma$,
for every $v_q\in\D$ we obtain
\beann
\sigma^*(i_\Gamma\wLD)(v_q) &=&
(i_\Gamma\wLD)_{\sigma(q)} (T_q\sigma(v_q)) =
{\wLD}_{\sigma(q)} (\Gamma(\sigma(q)),T_q\sigma(v_q))
\\ &=&
{\wLD}_{\sigma(q)} (T_q\sigma(\sigma(q)),T_q\sigma(v_q)) =
(\sigma^*\wLD)_q(\sigma(q),v_q) =
i_{\sigma}(\sigma^*\wLD)(v_q) ,
\eeann
and the result follows.
\qed 
\end{proof}

%%%%%%%%%%%%%%%%%%%%%%%%%%%%%%%%%%%%%%%%%%%%%%%%%%%%%%%%%%%%%%%%
\subsection{The restricted Hamilton--Jacobi problem}
%%%%%%%%%%%%%%%%%%%%%%%%%%%%%%%%%%%%%%%%%%%%%%%%%%%%%%%%%%%%%%%%

A solution of the 
\emph{(restricted) Lagrangian nonholonomic Hamilton--Jacobi problem}
is a solution $\sigma$ 
of the general Lagrangian nonholonomic Hamilton--Jacobi problem
which moreover satisfies the condition
$$
\sigma^*\wLD=0 .
$$
According to the preceding Proposition,
it follows that $\sigma$ must also satisfy $\sigma^*\eLD=0$.
Notice that, for $v\in\D$, we have 
\[
\langle \sigma^*\eLD,v\rangle=
\langle \sigma^*(dE_L\vert_{\TDD}),v\rangle=
\langle dE_L,\Tan \sigma(v)\rangle=
\langle d(\sigma^*E_L),v\rangle \ ,
\]
so that $\sigma^*\eLD=d(\sigma^*E_L)|_{\D}$.

Summarizing, we have proved the following:

\begin{proposition}
A section $\sigma\in\Sec{\D}$ 
is a solution to the Lagrangian nonholonomic Hamilton--Jacobi problem
if, and only if, $T\sigma(\D)\subset\TDD$
is a Lagrangian subbundle of $(\TDD,\wLD)$ 
and $d (\sigma^*E_L)\in\Sec{\Do}$.
\end{proposition}
Note that, when the distribution is bracket-generating,
the last condition means that the energy is constant,
as we have seen in Proposition~\ref{prop-brac}
at the end of the preceding section.

%%%%%%%%%%%%%%%%%%%%%%%%%%%%%%%%%%%%%%%%%%%%%%%%%%%%%%%%%%%%%%%%
%%%%%%%%%%%%%%%%%%%%%%%%%%%%%%%%%%%%%%%%%%%%%%%%%%%%%%%%%%%%%%%%
\section{Coordinate expressions and quasivelocities}
%%%%%%%%%%%%%%%%%%%%%%%%%%%%%%%%%%%%%%%%%%%%%%%%%%%%%%%%%%%%%%%%
%%%%%%%%%%%%%%%%%%%%%%%%%%%%%%%%%%%%%%%%%%%%%%%%%%%%%%%%%%%%%%%%

In order to find local expressions for the objects we have defined, 
we can use local coordinates in the base $Q$ and 
a set of linear coordinates (quasivelocities \cite{BMZ-qv,CNS-qv}) on the tangent bundle 
adapted to the distribution $\D$. 
This will greatly simplify many expressions.   
 
Let $(x^i)$ be local coordinates on $Q$ and choose 
a local basis $\{e_\alpha\}$ of sections of $\D$. 
Complete with $\{e_A\}$ to a local basis $\{e_\alpha,e_A\}$ of $\vf{Q}$, 
and denote the associated linear coordinates by $(y^\alpha,y^A)$, that is,
$y^\alpha = \widehat{e^\alpha}$ and $y^A = \widehat{e^A}$, 
where $\{e^\alpha,e^A\}$ is the dual basis. 
So we have coordinates $(x^i,y^\alpha,y^A)$ of $TQ$.
In these coordinates the constraints read $y^A=0$, 
so they are adapted to the submanifold $D \subset TQ$,
and $(x^i,y^\alpha)$ can be used as coordinates for~$\D$.

In the local coordinate system $(x^i)$ on $Q$, 
the elements of the basis $e_\alpha\in\vectorfields{Q}$ are given by
\[
e_\alpha=\rho^i_\alpha\pd{}{x^i},
\]
for some local functions $\rho^i_\alpha\in\cinfty{Q}$. 
The bracket of the sections $e_\alpha$ is of the form
\[
[e_\alpha,e_\beta]=C^\gamma_{\alpha\beta}e_\gamma+C^A_{\alpha\beta}e_A,
\]
with $C^i_{\alpha\beta}\in\cinfty{Q}$ local functions on $Q$. 
The constraints are integrable (holonomic) if, and only if, 
$C^A_{\alpha\beta}=0$. 
The vector fields $e_A$ have a similar expression, $e_A=\rho^i_A\pd{}{x^i}$.

\begin{remark}
In the classical literature, 
the functions of the type $C^a_{bc}$ are known as 
Hamel's transpositional symbols~\cite{Ha},
which obviously are nothing but the structure coefficients 
(in the Cartan's sense) 
of the moving frame $\{e_a\}$, see e.g.~\cite{EhKoMoRi}. 
Similar expressions arise in the theory of Lie algebroids, 
where the use of quasivelocities appears naturally
\cite{nonholoid, LSDLA, LMLA}.
\end{remark}

Associated with the above basis and coordinates 
we can find a local basis of sections of $\TDD$, that is, 
a family of $2r$ vector fields tangent to $\D$ which moreover 
project point-wise to vectors on $\D$. 
The coordinate vector fields 
$\partial/\partial x^i$ and $\partial/\partial y^\alpha$ 
are a basis of vector fields tangent to $\D$. 
The vector fields $\partial/\partial y^\alpha$ are vertical 
so that they project (through $T\tau$) to the zero section of $\D$. 
However, the vectors $\partial/\partial x^i$ 
do not (in general) project to admissible velocities. 
Taking an appropriate linear combination we have that 
the vector fields $\rho^i_\alpha \partial/\partial x^i$ 
project to $e_\alpha$, 
so that they are sections of $\TDD$. 
Moreover, since they are linearly independent, we have got a basis 
of sections of $\TDD$:
\[
\mathcal{X}_\alpha=\rho^i_\alpha\pd{}{x^i}
\quad\text{and}\quad
\mathcal{V}_\alpha=\pd{}{y^\alpha}.
\]
This basis of sections of $\TDD$ can be completed to 
a basis of sections of $T\D$ by adding the vector fields 
\[
\mathcal{X}_A=\rho^i_A\pd{}{x^i},
\]
and the brackets of the vector fields of such a basis are given by  
\[
[\mathcal{X}_\alpha,\mathcal{X}_\beta]=
C^\gamma_{\alpha\beta}\mathcal{X}_\gamma+C^A_{\alpha\beta}\mathcal{X}_A,
\quad
[\mathcal{X}_\alpha,\mathcal{V}_\beta]=0,
\quad
[\mathcal{V}_\alpha,\mathcal{V}_\beta]=0.
\]

A \sode\ $\Gamma\in\Sec{\TDD}$ is a vector field tangent to $\D$ 
and such that $T\tau(\Gamma(v))=v$ for every $v\in\D$. 
It follows that it is of the form 
\[
\Gamma=y^\alpha\,\mathcal{X}_\alpha+f^\alpha(x^i,y^\beta)\mathcal{V}_\alpha,
\]
for some local functions $f^\alpha\in\cinfty{\D}$. 
The differential equations for its integral curves are 
\[
\dot{x}^i=\rho^i_\alpha y^\alpha
\qquad
\dot{y}^\alpha=f^\alpha(x^i,y^\beta). 
\]

\bigskip
The above expressions can be specialized to the frequent case when 
the constraints are given by expressing
some velocities as linear functions of some other velocities, 
\[
\dot{x}^A-B^A_\alpha(x)\dot{x}^\alpha=0.
\]
The local basis $\{e_\alpha,e_A\}$ can be taken to be
\[
e_\alpha=\pd{}{x^\alpha}+B^A_\alpha\pd{}{x^A}\ ,
\qquad
e_A=\pd{}{x^A}\ ,
\]
and therefore the adequate quasivelocities are
\[
y_\alpha=\dot{x}_\alpha \,,
\qquad
y^A=\dot{x}^A-B^A_\alpha(x)\dot{x}^\alpha.
\]
The natural velocities are then given in terms of the quasivelocities by
\[
\dot{x}_\alpha=y_\alpha \,,
\qquad
\dot{x}^A=y^A+B^A_\alpha(x)y^\alpha.
\]
The local basis $\{\mathcal{X}_\alpha,\mathcal{V}_\alpha\}$ of sections of $\TDD$ can be given in terms of the natural coordinates by
\[
\mathcal{X}_\alpha =
\pd{}{x^\alpha}
+ B^A_\alpha\pd{}{x^A}
+ \dot{x}^\beta e_\alpha(B^A_\beta)\pd{}{\dot{x}^A} \,,
\qquad 
\mathcal{V}_\alpha =
\pd{}{\dot{x}^\alpha}
+ B^A_\alpha\pd{}{\dot{x}^A} \,,
\]
and we can complete it to a local basis of $\vf{\D}$ with 
\[
\mathcal{X}_A=\pd{}{x^A} \,.
\]
We can further complete to a local basis of $\vf{TQ}$ with 
\[
\mathcal{V}_A=\pd{}{\dot{x}^A} \,.
\]
The commutators of the elements of this basis of sections of $\TDD$ are
\[
[\mathcal{X}_\alpha,\mathcal{X}_\beta]=
R^A_{\alpha\beta}\left(\pd{}{x^A}+\pd{B^B_\gamma}{x^A}v^\gamma\pd{}{v^B}\right),
\quad
[\mathcal{X}_\alpha,\mathcal{V}_\beta]=0,
\quad
[\mathcal{V}_\alpha,\mathcal{V}_\beta]=0,
\]
where $R^A_{\alpha\beta}\equiv[e_\alpha,e_\beta]x^A=
e_\alpha(B^A_\beta)-e_\beta(B^A_\alpha)$.

%%%%%%%%%%%%%%%%%%%%%%%%%%%%%%%%%%%%%%%%%%%%%%%%%%%%%%%%%%%%%%%%
\paragraph{Lagrange--D'Alembert equations}

The constrained Lagrangian system $(L,\D)$ is regular if 
the restriction of the fibered Hessian $G^L$ to $\D$ 
is a regular bilinear tensor at every point. 
This restricted Hessian $\GLD$ has a particularly simple expression 
in the coordinates $(x^i,y^\alpha,y^A)$,
\[
\GLD[(x^i,y^\alpha)](e_\alpha,e_\beta)
=
\pd{^2L}{y^\alpha\partial y^\beta}(x^i,y^\alpha,0).
\]

The local expression of Lagrange--d'Alembert equations can be easily written 
in these coordinates without the need of Lagrange multipliers. 
By contracting the equations $i_\Gamma\wLD-\eLD=0$ with 
the elements of the basis $\{\mathcal{X}_\alpha,\mathcal{V}_\alpha\}$, 
and taking into account the constraints $y^A=0$, 
these equations read 
\begin{equation}
  \label{LD-vf}
    \Gamma\left(\pd{L}{y^\alpha}\right) 
     +\pd{L}{y^\gamma}C^\gamma_{\alpha\beta}y^\beta 
     -\rho^i_\alpha\pd{L}{x^i}=-\pd{L}{y^A}C^A_{\alpha\beta}y^\beta ,
\end{equation}
where $\Gamma=y^\alpha\mathcal{X}_\alpha+f^\alpha\mathcal{V}_\alpha$ 
is the \sode\ vector field we are looking for.

Taking into account the second-order condition,
the differential equations for the solutions of the dynamics are 
\begin{equation}
\label{LD-edo}
\begin{aligned}
&\dot{x}^i=\rho^i_\alpha y^\alpha,
\\
&\frac{d}{dt}\left(\pd{L}{y^\alpha}\right) 
 +\pd{L}{y^\gamma}C^\gamma_{\alpha\beta}y^\beta 
 -\rho^i_\alpha\pd{L}{x^i}=-\pd{L}{y^A}C^A_{\alpha\beta}y^\beta ,
\\
&y^A=0 \,.
\end{aligned}
\end{equation}

%%%%%%%%%%%%%%%%%%%%%%%%%%%%%%%%%%%%%%%%%%%%%%%%%%%%%%%%%%%%%%%%
\paragraph{General Hamilton--Jacobi problem}

By evaluating the equations~\eqref{LD-vf} on the image of the section $\sigma$,
\textit{i.e.}\ at a point of the form $y^\alpha=\sigma^\alpha(x)$, 
and taking into account the general Hamilton--Jacobi condition, 
Eq.~(\ref{genintrin}),
which can be expressed as a relation between differential operators as
$\sigma^* \circ \mathcal{L}_\Gamma = \mathcal{L}_\sigma \circ \sigma^*$, 
we obtain
\begin{equation}
\label{GHJ-local}
   \mathcal{L}_\sigma\left(\pd{L}{y^\alpha}\circ\sigma\right) 
  +\left(\pd{L}{y^\gamma}\circ\sigma\right)C^\gamma_{\alpha\beta}\sigma^\beta 
  -\rho^i_\alpha\left(\pd{L}{x^i}\circ\sigma\right)
 =-\left(\pd{L}{y^A}\circ\sigma\right)C^A_{\alpha\beta}\sigma^\beta,
\end{equation}
which is the local expression of the general Hamilton--Jacobi equation. 
In order to find the solutions of the dynamics 
these equations must be supplemented with the differential equations 
for the integral curves of $\sigma$, \textit{i.e.}\ 
$\dot{x}^i=\rho^i_\alpha\sigma^\alpha$.

With a simplified notation, the equations to be solved are
\begin{equation}
  \label{GHJ-edo}
  \begin{aligned}
    &\dot{x}^i=\rho^i_\alpha\sigma^\alpha,\\
    &\mathcal{L}_\sigma\left(\pd{L}{y^\alpha}\right) 
      +\pd{L}{y^\gamma}C^\gamma_{\alpha\beta}\sigma^\beta 
      -\rho^i_\alpha\pd{L}{x^i}=-\pd{L}{y^A}C^A_{\alpha\beta}\sigma^\beta,\\
  \end{aligned}
\end{equation}
where all the partial derivatives of the Lagrangian must be evaluated 
at points of the form $(x^i,\sigma^\alpha(x))$ 
before taking further derivatives.
We remark the formal similarity of these equations~\eqref{GHJ-edo} 
and equations~\eqref{LD-edo}, 
which are obtained formally by the substitution 
$y^\alpha=\sigma^\alpha(x)$ everywhere.

%%%%%%%%%%%%%%%%%%%%%%%%%%%%%%%%%%%%%%%%%%%%%%%%%%%%%%%%%%%%%%%%
\paragraph{Restricted Hamilton--Jacobi problem}

For the expression of the Hamilton--Jacobi equation 
we need the explicit expression of the form $\wLD$. 
This coordinate expression can be easily found by using the relation 
\[
\wLD(X,Y)=
\-d\theta_L(Y,X)=\mathcal{L}_Y(\theta_L(X))-\mathcal{L}_X(\theta_L(Y))+\theta_L([X,Y]),
\qquad 
X,Y\in\sec{\TDD},
\]
applied to the elements of the basis 
$\{\mathcal{X}_\alpha,\mathcal{V}_\alpha\}$. 
It turns out that
\begin{align*}
\wLD(\mathcal{X}_\alpha,\mathcal{X}_\beta)
&=
\rho^i_\beta\pd{^2L}{y^\alpha\partial x^i}
-\rho^i_\alpha\pd{^2L}{y^\beta\partial x^i}
+\pd{L}{y^\gamma}C^\gamma_{\alpha\beta}
+\pd{L}{y^A}C^A_{\alpha\beta} \,,
\\
\wLD(\mathcal{X}_\alpha,\mathcal{V}_\beta)
&=
\pd{^2L}{y^\alpha\partial y^\beta} \,,
\\
\wLD(\mathcal{V}_\alpha,\mathcal{V}_\beta)
&=
0 \,,
\end{align*}
where all the partial derivatives of the Lagrangian are taken 
at points in the constraint subbundle $\D$, 
\textit{i.e.}\ in the submanifold $y^A=0$. 
Therefore, the local expression of $\wLD$ is 
\[
\wLD=
\frac{1}{2}\left(
  \rho^i_\beta\pd{^2L}{y^\alpha\partial x^i}-\rho^i_\alpha\pd{^2L}{y^\beta\partial x^i}
  +\pd{L}{y^\gamma}C^\gamma_{\alpha\beta}+\pd{L}{y^A}C^A_{\alpha\beta}
\right) \mathcal{X}^\alpha\wedge\mathcal{X}^\beta
+
\pd{^2L}{y^\alpha\partial y^\beta}\mathcal{X}^\alpha\wedge\mathcal{V}^\beta ,
\]
where $\{\mathcal{X}^\alpha,\mathcal{V}^\alpha\}$ is 
the dual basis of $\{\mathcal{X}_\alpha,\mathcal{V}_\alpha\}$.
%\begin{small}
%An expression separating the part which depends only on the constrained Lagrangian $\ell(x^i,y^\alpha)=L(x^i,y^\alpha,0)$is the following 
%\begin{align*}
%\wLD=
%\frac{1}{2}\left(\rho^i_\beta\pd{^2\ell}{y^\alpha\partial x^i}-\rho^i_\alpha\pd{^2\ell}{y^\beta\partial x^i}+\pd{\ell}{y^\gamma}C^\gamma_{\alpha\beta}\right)\mathcal{X}^\alpha\wedge\mathcal{X}^\beta
%+\pd{^2\ell}{y^\alpha\partial y^\beta}\mathcal{X}^\alpha\wedge\mathcal{V}^\beta&+\\
%&+\frac{1}{2}J_AC^A_{\alpha\beta}\mathcal{X}^\alpha\wedge\mathcal{X}^\beta
%\end{align*}
%where $J_A=\pd{L}{y^A}(x^i,y^\alpha,0)$.
%\end{small}

To compute pullbacks we can proceed as follows. 
If $\map{\sigma}{Q}{D}$ is a section of $\D$ then 
the map $\CMcal{T}\sigma$ is determined by
\[
\CMcal{T}\sigma(e_\alpha) =
\mathcal{X}_\alpha+\rho^i_\alpha\pd{\sigma^\beta}{x^i}\mathcal{V}_\beta,
\] 
and therefore the pullback of the dual basis $\{\mathcal{X}^\alpha,\mathcal{V}^\alpha\}$ is 
\[
\sigma^*\mathcal{X}^\alpha=e^\alpha
\quad\text{and}\quad
\sigma^*\mathcal{V}^\alpha=\rho^i_\beta\pd{\sigma^\alpha}{x^i}e^\beta.
 \]
From this it is straightforward to calculate the local expression of $\sigma^*\wLD$:
\begin{align*}
%\[
\sigma^*\wLD
&=
\\ %%
\frac{1}{2}
&
\left[
 \rho^i_\beta \left(
   \pd{^2L}{y^\alpha\partial x^i}+\pd{^2L}{y^\alpha\partial y^\gamma}\pd{\sigma^\gamma}{x^i}
 \right)
-\rho^i_\alpha\left(
   \pd{^2L}{y^\beta\partial x^i}+\pd{^2L}{y^\beta\partial y^\gamma}\pd{\sigma^\gamma}{x^i}
 \right)
+\pd{L}{y^\gamma}C^\gamma_{\alpha\beta}
+\pd{L}{y^A}C^A_{\alpha\beta}
\right]
e^\alpha \!\wedge e^\beta.
% \]
\end{align*}
The equation $\sigma^*\wLD=0$ is equivalent to 
the vanishing of the expression between braces:
\begin{equation}
\label{lagrangeanity-local}
 \rho^i_\beta\left(\pd{^2L}{y^\alpha\partial x^i}+\pd{^2L}{y^\alpha\partial y^\gamma}\pd{\sigma^\gamma}{x^i}\right)
-\rho^i_\alpha\left(\pd{^2L}{y^\beta\partial x^i}+\pd{^2L}{y^\beta\partial y^\gamma}\pd{\sigma^\gamma}{x^i}\right)
+\pd{L}{y^\gamma}C^\gamma_{\alpha\beta}+\pd{L}{y^A}C^A_{\alpha\beta}=0 .
\end{equation}

Alternatively, one can calculate $\sigma^*\wLD(e_\alpha,e_\beta)=-d(\sigma^*\theta_L)(e_\alpha,e_\beta)$, from where the  vanishing of $\sigma^*\wLD=0$ is equivalent to the equations 
\[
 \mathcal{L}_{e_\alpha}\left(\pd{L}{y^\beta}\circ\sigma\right)
-\mathcal{L}_{e_\beta}\left(\pd{L}{y^\beta}\circ\sigma\right)
-\left(\pd{L}{y^\gamma}\circ\sigma\right)C^\gamma_{\alpha\beta}
=
\left(\pd{L}{y^A}\circ\sigma\right)C^A_{\alpha\beta}.
\]
%Or simply
%\[
% \mathcal{L}_{e_\alpha} P_\beta
%-\mathcal{L}_{e_\beta} P_\alpha
%-P_\gamma C^\gamma_{\alpha\beta}
%=
%J_A C^A_{\alpha\beta}.
%\]
%where we have written $P_\alpha=\pd{L}{y^\alpha}\circ\sigma$ and $J_A=\pd{L}{y^A}\circ\sigma$.

Finally, the pullback by $\sigma$ of the energy is easily calculated:
\[
\sigma^*E_L=\sigma^*\left(y^\alpha\pd{L}{y^\alpha}+y^A\pd{L}{y^A}-L\right)=
\sigma^\alpha\left(\pd{L}{y^\alpha}\circ\sigma\right)-L\circ\sigma.
\]

If the distribution is bracket-generating, 
then the equation $d(\sigma^*\eLD)\in\sec{\Do}$ can be substituted by
\[
\sigma^\alpha(x)\pd{L}{y^\alpha}(x^i,\sigma^\beta(x))
-L(x^i,\sigma^\beta(x) )
=
\mathrm{constant} \,.
\]

%%%%%%%%%%%%%%%%%%%%%%%%%%%%%%%%%%%%%%%%%%%%%%%%%%%%%%%%%%%%%%%%
%%%%%%%%%%%%%%%%%%%%%%%%%%%%%%%%%%%%%%%%%%%%%%%%%%%%%%%%%%%%%%%%
\section{Complete solutions}
%%%%%%%%%%%%%%%%%%%%%%%%%%%%%%%%%%%%%%%%%%%%%%%%%%%%%%%%%%%%%%%%
%%%%%%%%%%%%%%%%%%%%%%%%%%%%%%%%%%%%%%%%%%%%%%%%%%%%%%%%%%%%%%%%

The essential idea in the standard (unconstrained) Hamilton--Jacobi theory 
consists in finding a complete family of solutions to the problem 
(not only one particular solution).
In the present context a complete solution can be defined as follows:

\begin{definition}
Consider a solution $\sigma_\lambda$ to the general 
(respectively, restricted) 
Lagrangian nonholonomic Hamilton--Jacobi problem depending on
$r={\rm rank}\,(\D)$ additional parameters $\lambda\in\Lambda$
(where $\Lambda\subseteq\Real^r$ is some open set) 
and suppose that the map
$\Phi\colon Q\times\Lambda\to\D$ given by 
$\Phi(q,\lambda)=\sigma_\lambda(q)$
is a local diffeomorphism. 
In this case the family
$\{ \sigma_\lambda\,;\, \lambda\in\Lambda\}$
is said to be a {\rm complete solution} to the general 
(respectively, restricted) 
Lagrangian nonholonomic Hamilton--Jacobi problem.
\end{definition}

In other words, a \emph{complete solution} is a local diffeomorphism
 $\map{\Phi}{Q\times\Lambda}{\D}$ over the identity in $Q$,
 such that for every $\lambda\in\Lambda$ the section 
$\sigma_\lambda\in\Sec{\D}$
 given by $\sigma_\lambda(q)=\Phi(q,\lambda)$,
 is a solution to the general (restricted) Hamilton--Jacobi problem.

The interest of this notion is that all the integral curves of~$\Gamma$
 can be actually described as integral curves of appropriate vector fields in the
 complete solution. For every point $v\in{\rm Im}\,\Phi$ we take $q=\tau_Q(v)$
 and we find $\lambda\in\Real^r$ such that $\Phi(q,\lambda)=v$.
 The vector field $\sigma_\lambda$ is a solution to the generalized Hamilton--Jacobi
 problem, with $\sigma_\lambda(q)=v$. Taking the integral curve $\gamma(t)$ of
 $\sigma_\lambda$ passing through $q$, we have that $\dot{\gamma}(t)$ is the
 solution to the dynamics starting at $v$.

In what follows, for simplicity, we will assume that $\Phi\colon Q\times\Lambda\to\D$
is a global diffeomorphism. We then define the map $\map{F}{\D}{\Real^r}$ by
$F={\rm pr}_2\circ \Phi^{-1}$, where ${\rm pr}_2:Q\times\Lambda\to\Lambda$
is the projection onto the second factor. 

From the very definition, it follows that a complete solution provides the manifold
$\D$ with a foliation transverse to the fibers of $\tau \colon \D \to Q$,
the leaves being the image of the vector fields $\sigma_\lambda$,
and that the solution vector field $\Gamma$ is tangent to the leaves.
We now study this foliation with more detail, specially in the case of a complete
solution to the restricted problem.

\begin{proposition}
\label{foliation}
The following properties hold.
\begin{enumerate}
\item 
For every $\lambda\in\Lambda$ we have
 $F^{-1}(\lambda)={\rm Im}\,(\sigma_\lambda)$.
\item 
The map $T\Phi\colon TQ\times T\Lambda\to T\D$
restricts to a map
$\map{\CMcal{T}\Phi}{\D\times T\Lambda}{\TDD}$.
Moreover $\CMcal{T}\Phi$
is a diffeomorphism 
(a local diffeomorphism if $\Phi$ is a local diffeomorphism).
\item
The section 
$\bar{\Gamma} \in \Sec{\D \times T \Lambda \to Q \times \Lambda}$
defined by 
$\bar{\Gamma} = (\CMcal{T}\Phi)^{-1} \circ \Gamma \circ \Phi$
has the form  $\bar{\Gamma}(q,\lambda)=(\sigma_\lambda(q),0_\lambda)$.
\item
The components of the map $F$ are constants of motion.
\item
If $\Phi$ is a complete solution to the \emph{restricted} problem,
then the subbundles\\
$\{(v,0)\in\D\times T\Lambda\}$
and $\{(0,z)\in\D\times T\Lambda\}$
are Lagrangian subbundles of the symplectic bundle
 $\bigl(\D\times T\Lambda,\Phi^*(\wLD)\bigr)$.
\end{enumerate}
\end{proposition}
\begin{proof}
We will make use of the following fact:
\begin{equation}
\CMcal{T}\Phi(v_q,0_\lambda)=T\sigma_\lambda(v_q).
\label{star}
\end{equation}
Indeed, if $\gamma(s)$ is a curve in $Q$ such that $\dot{\gamma}(0)=v_q$, 
then
\begin{align*}
T\Phi(v_q,0_\lambda)f
&=(v_q,0_\lambda)(f\circ\Phi)
=\frac{d}{ds}f(\Phi(\gamma(s),\lambda))\at{s=0}\\
&=\frac{d}{ds}f(\sigma_\lambda(\gamma(s)))\at{s=0}
=\frac{d}{ds}(\sigma_\lambda^*f)(\gamma(s)))\at{s=0}\\
&=v_q(\sigma_\lambda^*f)
=T\sigma_\lambda(v_q)f
\end{align*}
for every function $f\in\cinfty{\D}$, which proves the equality (\ref{star}).
\\
Now let us proceed with the proof of the proposition:
\begin{enumerate}
\item
 $v\in F^{-1}(\lambda)$ $\Longleftrightarrow$ $F(v)=\lambda$      
   $\Longleftrightarrow$
 $\Phi^{-1}(v)=(\tau(v),\lambda)$
   $\Longleftrightarrow$
 $v=\Phi(\tau(v),\lambda)$ 
   $\Longleftrightarrow$
 $v=\sigma_\lambda(\tau(v))$ 
   $\Longleftrightarrow$ 
 $v\in{\rm Im}\,(\sigma_\lambda)$.
\item For every $(v_q,z_\lambda)\in TQ\times T\Lambda$
\[
T\tau(T\Phi(v_q,z_\lambda))=T(\tau\circ\Phi)(v_q,z_\lambda)=
T{\rm pr}_1(v_q,z_\lambda)=v_q \ .
\]
So, $T\Phi(v_q,z_\lambda)$ belongs to $\TDD$ if, and only if, $v_q \in \D$.
\item
We have just to prove that 
$\CMcal{T}\Phi(\sigma_\lambda(q),0_\lambda)=\Gamma(\Phi(q,\lambda))$,
 for every $(q,\lambda)\in Q\times\Lambda$. Using (\ref{star}), for
 $v_q=\sigma_\lambda(q)$ we have that
 $\CMcal{T}\Phi(\sigma_\lambda(q),0_\lambda)=
 T\sigma_\lambda(\sigma_\lambda(q))$, and taking into account that
 $T\sigma_\lambda\circ \sigma_\lambda=\Gamma\circ \sigma_\lambda$,
 we finally get
\[
\CMcal{T}\Phi(\sigma_\lambda(q),0_\lambda)
=T\sigma_\lambda(\sigma_\lambda(q))
=\Gamma(\sigma_\lambda(q))
=\Gamma(\Phi(q,\lambda)).
\]
\item
$F$ is constant on ${\rm Im}\,\sigma_\lambda$, and $\Gamma$ is tangent to
 ${\rm Im}\,\sigma_\lambda$, so the result follows.
\item
First, using (\ref{star}) we have
\begin{align*}
(\Phi^*\wLD)_{(q,\lambda)}\bigl((v_q,0_\lambda),(w_q,0_\lambda)\bigr)
&=\wLD[\Phi(q,\lambda)]\bigl(T\Phi(v_q,0_\lambda),T\Phi(w_q,0_\lambda)\bigr)\\
&=\wLD[\Phi(q,\lambda)]\bigl(T\sigma_\lambda(v_q),T\sigma_\lambda(w_q)\bigr)\\
&=(\sigma_\lambda^*\wLD)_q(v_q,w_q)=0
\end{align*}
Furthermore, notice that $\CMcal{T}\Phi(\Ver{{\rm pr}_1})=\Ver{\tau}$, 
which can be easily proved. Thus
$$
(\Phi^*\wLD)_{(q,\lambda)}\bigl((0_q,y_\lambda),(0_q,z_\lambda)\bigr)
=\wLD{}_{\Phi(q,\lambda)}\bigl(T\Phi(0_q,y_\lambda),T\Phi(0_q,z_\lambda)\bigr)
=0
$$
because $\Ver{\tau}$ is an isotropic (in fact Lagrangian) subbundle of the symplectic
 bundle~$(\TDD,\wLD)$.
\end{enumerate}
This finishes the proof.
\qed
 \end{proof}

In the case of a complete solution to the restricted problem,
 only the terms of the form $\Phi^*\wLD( (v,0),(0,z))$ can possibly be nonzero, 
and they can be expresed in terms of the Hessian,
\[
(\Phi^*\wLD)_{(q,\lambda)}( (v_q,0_\lambda),(0_q,z_\lambda))
=\GLD[\sigma_\lambda(q)](v_q,\bar{z}(q,\lambda)),
\]
where $\bar{z}$ is defined by
 $\xi^V(0_q,\bar{z}(q,\lambda))=T\Phi(0_q,z_\lambda)$. 
Indeed, by the definition of $\GLD$ we have
\begin{align*}
(\Phi^*\wLD)_{(q,\lambda)}\bigl((v_q,0_\lambda),(0_q,z_\lambda)\bigr)
&=\wLD[\Phi(q,\lambda)]\bigl(T\sigma_\lambda(v_q),T\Phi,(0_q,z_\lambda)\bigr)\\
&=\wLD[\sigma_\lambda(q)]
\bigl(T\sigma_\lambda(v_q),\xi^V(0_q,\bar{z}(q,\lambda))\bigr)\\
&=\GLD[\sigma_\lambda(q)](v,\bar{z}(q,\lambda)).
\end{align*}

%%%%%%%%%%%%%%%%%%%%%%%%%%%%%%%%%%%%%%%%%%%%%%%%%%%%%%%%%%%%%%%%
\paragraph{The nonholonomic bracket}

For a complete solution the map $F$ is a constant of the motion, that is,
 if we denote by $f_1,\ldots,f_r$ the components of $F$,
 then every function $f_i$ is a constant of the motion for  $\Gamma$.
 Conversely, a family of functionally independent first integrals
 $f_1,\cdots,f_r\in\cinfty{\D}$, satisfying the transversality condition
 $\det\bigl[\langle{df_\alpha}\,,\,{e_\beta^V}\rangle\bigr]\not= 0$
 (where $\{e_\alpha\}$ is any local basis for $\D$),  
defines a complete integral by
 means of $\Phi^{-1}(v)=\bigl(\tau(v),\bigl(f_1(v),\ldots,f_r(v)\bigr)\bigr)$. 

We now show that this functions are in involution with respect to the nonholonomic
 bracket. One of the possible constructions of such bracket is as follows. Given a
 function $g\in\cinfty{\D}$, we consider the section $\delta g\in\Sec{(\TDD)^*}$
 as the restriction of the differential of $g$ to $\TDD$; that is,
 $\delta g=dg|_{\TDD}$\,. Since the constrained system is regular,
 we can define the nonholonomic Hamiltonian section $\eta_g\in\Sec{\TDD}$
 by means $i_{\eta_g}\wLD=\delta g$.
 Then we define the nonholonomic bracket of two functions
 $f,g\in\cinfty{\D}$ by means of $\nhbr{f}{g}=\wLD(\eta_f,\eta_g)$.
 This bracket is skewsymmetric but it does not satisfy the Jacobi identity,
 except if the constraints are actually holonomic.

\begin{theorem}
If $F = (f_1,f_2,\ldots,f_r)$ then $\nhbr{f_i}{f_j}=0$.
\end{theorem}
\begin{proof} 
We will show that the sections $\eta_{f_i}$ are of the form
 $\eta_{f_i}=\CMcal{T}\Phi(X_f,0)$. Indeed, let
 $Z\in\Sec{\D\times T\Lambda\to Q\times\Lambda}$
 be the section such that $\CMcal{T}\Phi\circ Z_i=\eta_{f_i}\circ\Phi$.
 For every $v\in\D$, let $q=\tau(v)$ and $\lambda=F(v)$,
 so that $\Phi(q,\lambda)=v$. For every $w\in\D[q]$ we have
\begin{align*}
(\Phi^*\wLD)_v\bigl(Z_i(v),(w,0)\bigr)
&=\wLD[v]\bigl(T\Phi(Z_i(v)),T\Phi(w,0)\bigr)
=\wLD[v]\bigl(\eta_{f_i}(v),T\sigma_\lambda(w)\bigr)\\
&=T\sigma_\lambda(w)\cdot {f_i}
=w\cdot(f_i\circ \sigma_\lambda)
=w\cdot(\lambda_i)=0
\end{align*}
where we have used that $F\circ\sigma_\lambda=\lambda$ (constant),
and hence $f_i\circ\sigma_\lambda=\lambda_i$.
Therefore $Z_i$ takes values in the orthogonal with respect to $\Phi^*\wLD$
of the subbundle $\{(v,0)\in\D\times T\Lambda\}$.
Since this subbundle is Lagrangian, we have that $Z_i$ takes values on it,
\textit{i.e.} it is of the form $Z_i=T\Phi(W_i,0)$. 
But then 
\[
\nhbr{f_i}{f_j}
=\wLD(\eta_{f_i},\eta_{f_j})
=\wLD\bigl(T\Phi(W_i,0),T\Phi(W_j,0)\bigr)
=(\Phi^*\wLD)\bigl((W_i,0),(W_j,0)\bigr)
=0,
\]
which finishes the proof.
\qed
\end{proof}

%%%%%%%%%%%%%%%%%%%%%%%%%%%%%%%%%%%%%%%%%%%%%%%%%%%%%%%%%%%%%%%%
%%%%%%%%%%%%%%%%%%%%%%%%%%%%%%%%%%%%%%%%%%%%%%%%%%%%%%%%%%%%%%%%
\section{Example}
%%%%%%%%%%%%%%%%%%%%%%%%%%%%%%%%%%%%%%%%%%%%%%%%%%%%%%%%%%%%%%%%
%%%%%%%%%%%%%%%%%%%%%%%%%%%%%%%%%%%%%%%%%%%%%%%%%%%%%%%%%%%%%%%%

%%%%%%%%%%%%%%%%%%%%%%%%%%%%%%%%%%%%%%%%%%%%%%%%%%%%%%%%%%%%%%%%
\subsection{The nonholonomic free particle}
%%%%%%%%%%%%%%%%%%%%%%%%%%%%%%%%%%%%%%%%%%%%%%%%%%%%%%%%%%%%%%%%

Every one-dimensional distribution is integrable, so that the easiest example of a
nonholonomic system is obtained in $\Real^3$ by a 2-dimensional distribution.
By an adequate change of coordinates, the annihilator of $\D$ is generated by the
1-form $dx_3-x_2dx_1$. The following example consists on a free particle 
under the action of such a constraint, and it is known as the nonholonomic
free particle~\cite{BS-nonh,GM-nh,Ros}.

Consider a particle moving in $Q = \Real^3$, with Lagrangian function
\[
L=\frac{1}{2}(\dot{x}_1^2+\dot{x}_2^2+\dot{x}_3^2) .
\]
We have
$\omega_L =
\d x_1 \wedge \d \dot{x}_1+
\d x_2 \wedge \d \dot{x}_2+
\d x_3 \wedge \d \dot{x}_3$
and
$\d E_L =
\dot{x}_1\d\dot{x}_1+\dot{x}_2\d\dot{x}_2+\dot{x}_3\d\dot{x}_3$,
so the unconstrained dynamics is the well-known
 free dynamics described by the vector field
\[
\Gamma_0 =
\widehat\omega^{-1}_L \circ \d E_L =
\dot{x}_1 \tanvec{x_1} + \dot{x}_2 \tanvec{x_2} + \dot{x}_3 \tanvec{x_3} .
\]

We introduce the nonholonomic constraint
\[
\phi=\dot{x}_3-x_2\dot{x}_1 = 0,
\]
so that the constraint submanifold is
$\D =
\{ (x_1,x_2,x_3;\dot{x}_1,\dot{x}_2,\dot{x}_3) \in \Tan Q \mid \dot{x}_3 = x_2\dot{x}_1 \}$.
Applying D'Alembert's principle for nonholonomic dynamics we get
\[
\Gamma=
\restric{
\left(
\dot{x}_1\tanvec{x_1}+\dot{x}_2\tanvec{x_2}+x_2\dot{x}_1\tanvec{x_3}-
\frac{x_2\dot{x}_2\dot{x}_1}{x_2^2+1}\tanvec{\dot{x}_1}+
\frac{\dot{x_2}\dot{x}_1}{x_2^2+1}\tanvec{\dot{x}_3}
\right)}{\D}.
\]

As a basis $\{e_\alpha\}$ of sections of $\D$ we can take,
\[
e_1=\pd{}{x_1}+x_2\pd{}{x_3} \,, \quad
e_2=\pd{}{x_2} \,,
\]
which we can complete with the vector field
\[
e_3=\pd{}{x_3} \,.
\]
The associated quasivelocities are related to the velocities by 
\begin{equation}
\label{quasivelocities-ex}
\begin{aligned}
y_1&=\dot{x}_1 \\                
y_2&=\dot{x}_2 \\              
y_3&=\dot{x}_3-x_2\dot{x}_1    
\end{aligned}
\qquad\qquad
\begin{aligned}
\dot{x}_1&=y_1\\
\dot{x}_2&=y_2\\
\dot{x}_3&=y_3+x_2y_1.\\
\end{aligned}
\end{equation}
The corresponding basis $\{\mathcal{X}_\alpha,\mathcal{V}_\alpha\}$ 
can be expressed in terms of the natural coordinates on the tangent bundle as 
\begin{align*}
&\mathcal{X}_1=\pd{}{x_1}+x_2\pd{}{x_3}
&&\mathcal{X}_2=\pd{}{x_2}+\dot{x_1}\pd{}{\dot{x}_3}
\\
&\mathcal{V}_1=\pd{}{\dot{x}_1}+x_2\pd{}{\dot{x}_3}
&&\mathcal{V}_2=\pd{}{\dot{x}_2},
\end{align*}
and it is completed to a basis of sections of $T\D$ with the vector field
\[
\mathcal{X}_3=\pd{}{x_3} .
\]
We have that the symplectic section is given by
\[
\wLD=x_2y_1\,\mathcal{X}^1\wedge\mathcal{X}^2+(1+x_2^2)\,\mathcal{X}^1
\wedge\mathcal{V}^1+\mathcal{X}^2\wedge\mathcal{V}^2,
\]
and the 1-form $\eLD$ is 
\[
\eLD=x_2y_1^2\,\mathcal{X}^2+(1+x_2^2)y_1\,\mathcal{V}^1+y_2\,\mathcal{V}^2.
\]
From here, the dynamical section $\Gamma$, such that $i_\Gamma\wLD=\eLD$, 
is 
\[
\Gamma=y_1\mathcal{X}_2+
y_2\mathcal{X}_2-\frac{x_2}{1+x_2^2}y_1y_2\mathcal{V}_1,
\]
and the integral curves of $\Gamma$ are the solutions to
\begin{equation}
\label{LD-ex}
\dot{x}_1=y_1\, ,
\qquad
\dot{x}_2=y_2\, ,
\qquad
\dot{y}_1=-\frac{x_2}{1+x_2^2}y_1y_2\, ,
\qquad
\dot{y}_2=0\, ,
\end{equation}
together with the constraint $\dot{x}_3=x_2\dot{x}_1$.

Note that in this example $\D$ is a bracket-generating distribution. 
Indeed, $e_1$, $e_2$ and 
\[
[e_1,e_2] =
\frac{1}{(x_2^2+1)^{3/2}} \left( \tanvec{x_3}-x_2\tanvec{x_1} \right)
\]
are linearly independent at each point of $\Real^3$. 
It follows that there is only one orbit for this distribution, 
the full space $\Real^3$, 
and hence any pair of points can be joined by 
concatenation of integral curves of vector fields belonging 
to the distribution $\D$.

%%%%%%%%%%%%%%%%%%%%%%%%%%%%%%%%%%%%%%%%%%%%%%%%%%%%%%%%%%%%%%%%
\subsection{The Hamilton--Jacobi problem}
%%%%%%%%%%%%%%%%%%%%%%%%%%%%%%%%%%%%%%%%%%%%%%%%%%%%%%%%%%%%%%%%

Let us state the Hamilton--Jacobi problem for this dynamics.
According to the general discussion,
we wish to find the vector fields $X$ in~$Q$ such that:
\begin{enumerate}
\item
$X$ takes values in~$\D$, and

\item
$X$ and $\Gamma$ are $X$-related:
$\Tan X \circ X = \Gamma \circ X$.
\end{enumerate}
From the first condition we have that $X$ has the form
\begin{equation}
X = f\left( \tanvec{x_1}+x_2\tanvec{x_3}\right) +  g\tanvec{x_2} \ ,
\label{Xex}
\end{equation}
and the second condition leads to
\[
\Lie_X f = -\frac{x_2}{x_2^2+1}fg \ , \quad
\Lie_X g = 0 \, ,
\]
or, more explicitly,
\begin{equation}
f\derpar{f}{x_1}+g\derpar{f}{x_2}+x_2f\derpar{f}{x_3}=-fg \frac{x_2}{x_2^2+1}
 \ , \quad
f\derpar{g}{x_1} + g\derpar{g}{x_2} + x_2f \derpar{g}{x_3} = 0 \, .
\label{conds}
\end{equation}

It is easy to find particular solutions of these equations. 
For instance, $f=1$, $g=0$
is a solution and also $f=0$, $g=1$ is a solution. From them we can find
some integral curves of the dynamics $\Gamma$. However, to obtain
all the integral curves of the dynamical vector field we need to look for
a complete solution.

\begin{remark}
An easier way to find these equations 
(together with the equations for the integral curves of $X$) 
is from D'Alembert equations~\eqref{LD-ex} 
by the substitution $y_1=f$ and $y_2=g$. 
We get the equations 
\[
\dot{x}_1=f,\quad
\dot{x}_2=g,\quad
\dot{f}=-\frac{x_2}{1+x_2^2}fg,\quad\text{and}\quad
\dot{g}=0.
\]
Equations~\eqref{conds} follow from this when expanding 
the total time derivatives of the functions $f$ and $g$ 
and using the first two equations.
\end{remark}

%%%%%%%%%%%%%%%%%%%%%%%%%%%%%%%%%%%%%%%%%%%%%%%%%%%%%%%%%%%%%%%%
\subsection{A complete solution}
%%%%%%%%%%%%%%%%%%%%%%%%%%%%%%%%%%%%%%%%%%%%%%%%%%%%%%%%%%%%%%%%

In order to get a complete solution, 
we look for a diffeomorphism $\Phi \colon Q \times \Lambda \to \D$, 
with $\Lambda = \R^2$, such that $\Phi(q,\lambda)=X_\lambda(q)$. 
In our case, taking (\ref{Xex}) into account, 
this means $\Phi(x_1,x_2,x_3;\lambda_1,\lambda_2)=(x_1,x_2,x_3;f,g,x_2f)$,  
where the functions $f,g$ satisfy (\ref{conds}).

As we already know, any solution to the free problem with values in $\D$ 
is also a solution to the constrained problem. 
There are some obvious solutions of the free problem 
that are constant vector fields, one of which,
$X=(0,\mathrm{constant},0)$, takes values in~$\D$. 
Thus, as we look for a particular complete solution, we can choose
$g=\lambda_1$ (constant) and try to find a corresponding value of $f$ satisfying
\[
f\derpar{f}{x_1}+\lambda_1\derpar{f}{x_2}+x_2f\derpar{f}{x_3}=
-f\lambda_1 \frac{x_2}{x_2^2+1}
\]
If we assume that $f$ depends only on $x_2$, $f=f(x_2)$, then the above equation is
\[
\frac{df}{dx_2}=-\frac{x_2}{x_2^2+1}f ,
\]
whose solution is $f=\lambda_2/\sqrt{x_2^2+1}$. 
Hence we have a complete solution that can be expressed as
\begin{equation}
X_{\lambda_1,\lambda_2} =
\lambda_1 \tanvec{x_2}+\lambda_2\frac{1}{\sqrt{x_2^2+1}}\left(\tanvec{x_1}+
x_2\tanvec{x_3} \right)=
\lambda_1  X_1 + \lambda_2 X_2 ,
\label{equis1}
\end{equation}
with
$$
X_{1} = \tanvec{x_2} ,
\quad
X_{2} = \frac{1}{\sqrt{x_2^2+1}} \left( \tanvec{x_1}+x_2\tanvec{x_3} \right) \, .
$$
In other words, we have the diffeomorphism given by
$$
\Phi(x_1,x_2,x_3;\lambda_1,\lambda_2) 
= 
\left(x_1,x_2,x_3;\frac{\lambda_2}{\sqrt{x_2^2+1}},\lambda_1,
\frac{x_2\lambda_2}{\sqrt{x_2^2+1}} \right)
\equiv 
(x_1,x_2,x_3;\dot{x}_1,\dot{x}_2,\dot{x}_3) \ .
$$
From here we get
$$
\Phi^{-1}(x_1,x_2,x_3;\dot{x}_1,\dot{x}_2,\dot{x}_3)=
\left(x_1,x_2,x_3;\dot{x}_1\sqrt{x_2^2+1},\dot{x}_2\right)
\equiv (x_1,x_2,x_3;\lambda_1,\lambda_2)
$$
and, therefore, we obtain the following constants of motion
$$
f_1=\dot{x}_1 \sqrt{x_2^2+1} \, , \ f_2=\dot{x}_2 \,.
$$
Let us remark the linear expression of $X_{\lambda_1,\lambda_2}$,
 which is related to the fact that the conserved quantities are linear 
(in the velocities).

A straightforward calculation  shows that the solution
 that we have found is a solution to the restricted Hamilton--Jacobi problem,
 that is $X_{\lambda_1\lambda_2}^*\wLD=0$. Alternatively,
 we can calculate the pullback of $\omega_L$ 
\[
X^*(\omega_L) =
\frac{\lambda_2}{(x_2^2+1)^{3/2}} (dx_3-x_2 dx_1 ) \wedge dx_2 ,
\]
which is in the exterior ideal generated by $\Do$.

\medskip

The flow of $X_{\lambda_1,\lambda_2}$ can also be easily computed:
 when $\lambda_1 \neq 0$, its integral curves are
\beann
x_1(t) &=&
x_1^0 + \frac{\lambda_2}{\lambda_1}\left(\arg\sinh(x_2^0+\lambda_1t) -
 \arg\sinh(x_2^0) \right) ,
\\
x_2(t) &=&
x_2^0+\lambda_1t ,
\\
x_3(t) &=&
x_3^0 + \frac{\lambda_2}{\lambda_1}\left(\strut\sqrt{1+(x_2^0+\lambda_1t)^2} -
 \sqrt{1+(x_2^0)^2} \right) ;
\eeann
when $\lambda_1 = 0$, the expression of the flow is
\beann
x_1(t) &=&
x_1^0 + \frac{\lambda_2}{\sqrt{1+(x_2^0)^2}}t ,
\\
x_2(t) &=&
x_2^0 ,
\\
x_3(t) &=&
x_3^0 + \frac {\lambda_2 x_2^0}{\sqrt{1+(x_2^0)^2}}t .
\eeann
It follows that (the tangent lift of) these curves 
are the solutions of the nonholonomic problem.

%%%%%%%%%%%%%%%%%%%%%
%\bigskip

%----------------------------------------------------
%\\
%Note that in this example $\D \subset \Tan Q$ is 
%a nonintegrable tangent subbundle.
% Since $X_1$ and $X_2$ are linearly independent and take values in~$\D$,
% they span this subbundle and therefore they cannot be in involution:
%$$
%[X_1,X_2] =
%\frac{1}{(y^2+1)^{3/2}} \left( \tanvec{z}-y\tanvec{x} \right)\ .
%$$
%Note also that the three vector fields $X_1$, $X_2$ and $[X_1,X_2]$
% are linearly independent at each point.  

%The usage of the flow of a single vector field of these does not allow
% to move outside from a surface in~$Q$.

%The usage of the flows of some (up to 4) of them allows,
% from a given initial point $(x_0,y_0,z_0)$, to reach every other point.
%\\
%----------------------------------------------------
%%%%%%%%%%%%%%%%%%%%%%%%%%

%%%%%%%%%%%%%%%%%%%%%%%%%%%%%%%%%%%%%%%%%%%%%%%%%%%%%%%%%%%%%%%%
\paragraph*{Another complete solution}

We can obtain another complete solution by choosing 
$g=\lambda_1$ (constant), 
as above, but now we try $f=f(x_2,x_3$). 
Then the second equations of (\ref{conds}) holds, and the first one reads
\[
\lambda_1\derpar{f}{x_2}+x_2f\derpar{f}{x_3}=-f\lambda_1 \frac{x_2}{x_2^2+1},	
\]
which has a solution $\ds f=\frac{x_3\lambda_1-\lambda_2}{x_2^2+1}$.
 Hence we obtain another complete solution to the Hamilton--Jacobi problem:
\begin{equation}
X_{\lambda_1,\lambda_2}=
\frac{\lambda_1x_3-\lambda_2}{x_2^2+1}\tanvec{x_1}+
 \lambda_1\tanvec{x_2}+x_2\frac{\lambda_1x_3-\lambda_2}{x_2^2+1}\tanvec{x_3} =
 \lambda_1 X_1 + \lambda_2 X_2 ,
\label{equis}
\end{equation}
with
\begin{equation}
X_{1} =
\frac{x_3}{x_2^2+1} \left(\tanvec{x_1} + x_2\tanvec{x_3}\right)+ \tanvec{x_2}  ,
\quad
X_{2} = \frac{-1}{x_2^2+1} \left( \tanvec{x_1}+x_2\tanvec{x_3} \right) .
\end{equation}
This solution leads to the following constants of motion:
$$
f_1=x_3\dot x_2-\dot{x}_1(x_2^2+1) \, , \quad \dot f_2=x_2 \,.
$$
In this case the solution that we found is not a solution to the restricted problem,
 that is $X_{\lambda_1,\lambda_2}^*\wLD\neq0$. In fact, we have
\[
X^*(\omega_L) =
-2x_2 \frac{\lambda_1x_3-\lambda_2}{(x_2^2+1)^2} dx_1 \wedge dx_2 +
\frac{\lambda_1}{x_2^2+1} dx_1 \wedge dx_3 +
(\lambda_1x_3-\lambda_2)\frac{x_2^2-1}{(x_2^2+1)^2} dx_2 \wedge dx_3 .
\]
and hence 
\[
X^*\wLD =
-\frac{x_2}{1+x_2^2} \, (\lambda_1x_3-\lambda_2) \, e^1 \wedge e^2.
\]

%%%%%%%%%%%%%%%%%%%%%%%%%%%%%%%%%%%%%%%%%%%%%%%%%%%%%%%%%%%%%%%%
%%%%%%%%%%%%%%%%%%%%%%%%%%%%%%%%%%%%%%%%%%%%%%%%%%%%%%%%%%%%%%%%
% \section{Conclusions and outlook}
%%%%%%%%%%%%%%%%%%%%%%%%%%%%%%%%%%%%%%%%%%%%%%%%%%%%%%%%%%%%%%%%
%%%%%%%%%%%%%%%%%%%%%%%%%%%%%%%%%%%%%%%%%%%%%%%%%%%%%%%%%%%%%%%%

%%%%%%%%%%%%%%%%%%%%%%%%%%%%%%%%%%%%%%%%%%%%%%%%%%%%%%%%%%%%%%%%
\subsection*{Acknowledgments}
%%%%%%%%%%%%%%%%%%%%%%%%%%%%%%%%%%%%%%%%%%%%%%%%%%%%%%%%%%%%%%%%

We acknowledge the partial financial support of the
\emph{Ministerio de Educaci\'on y Ciencia}, projects MTM2005-04947, 
MEC-DGI MTM2006-10531, MTM2008-00689/MTM and MTM2008-03606-E/MTM,
FPA-2003-02948 and CO2-399.
XG also acknowledges financial support from the mobility program 
(ref.\ PR2008-0317) of the {\it Ministerio de Ciencia e Innovaci\'on}.
% We thank Mr.\ Jeff Palmer for his assistance in preparing the
%English version of the manuscript.

%%%%%%%%%%%%%%%%%%%%%%%%%%%%%%%%%%%%%%%%%%%%%%%%%%%%%%%%%%%%%%%%
%%%%%%%%%%%%%%%%%%%%%%%%%%%%%%%%%%%%%%%%%%%%%%%%%%%%%%%%%%%%%%%%

%%%%%%%%%%%%%%%%%%%%%%%%%%%%%%%%%%%%%%%%%%%%%%%%%%%%%%%%%%%%%%%%
%%%%%%%%%%%%%%%%%%%%%%%%%%%%%%%%%%%%%%%%%%%%%%%%%%%%%%%%%%%%%%%%

%%%%%%%%%%%%%%%%%%%%%%%%%%%%%%%%%%%%%%%%%%%%%%%%%%%%%%%%%%%%%%%%
%%%%%%%%%%%%%%%%%%%%%%%%%%%%%%%%%%%%%%%%%%%%%%%%%%%%%%%%%%%%%%%%
\end{document}